\documentclass[a4paper,times,10pt]{article}
\usepackage{a4wide}
\usepackage{graphicx}
\usepackage{amssymb}
\usepackage[super,square]{natbib}
\begin{document}
\baselineskip=22pt

\title{Tailoring Nano-Structures Using Copolymer Nanoimprint Lithography}

\author{Pascal Th\'ebault$^{1}$, Stefan Niedermayer$^{1}$, Stefan Landis$^{2}$, Nicolas Chaix$^{2}$,\\
Patrick Guenoun$^{1}$, Jean Daillant$^{1}$,
Xingkun Man$^{3}$, David Andelman$^{3}$,\\ Henri Orland$^{4}$\\
$^1$CEA, IRAMIS, SIS2M LIONS, CNRS, UMR $\rm n^{\circ}$ 3299, F-91191 Gif-sur-Yvette Cedex,\\ France.\\
$^2$CEA-LETI, F-38054 Grenoble Cedex, France\\
$^3$Raymond and Beverly Sackler School of Physics and Astronomy, \\Tel Aviv University, Ramat Aviv, Tel Aviv 69978, Israel\\
$^4$Institut de Physique Th\'eorique, CEA-Saclay, 91191 Gif-sur-Yvette Cedex,\\ France}

\maketitle

Microphase separation of thin films of block copolymers (BCPs) leads to tunable and perfect structures  at the block scale\cite{bates}.
However, because an abundance of defects is ever-present at larger scales~\cite{kellogg},
great efforts have been devoted to produce defect-free structures applicable to devices~\cite{bang, ruiz07a, stoyko05, cheng}.

The challenge of finding efficient ways to organize BCP films has been addressed by combining self-assembly at the nanometric scale
with top-down approaches used to facilitate large-scale film organization~\cite{bang}.
Among other techniques, this has been achieved by modulating the substrate energy using
e-beam lithography pre-patterning~\cite{stoyko05, cheng} or using a BCP underlayer\cite{ruiz07a}.
The epitaxy onto the  surface periodic structure guides BCP lamellae along patterns
such as bends or curves\cite{stoyko05,cheng}.
However, epitaxy requires the same spatial resolution as that of the expected device.

Graphoepitaxy is another technique where a substrate topography is created
in order to control the BCP domain orientation.
Because of topographical constraints, BCP ordered regions are obtained over micrometers\cite{ruiz07b,segalman01},
but the process has to be repeated for each new sample.

\begin{figure}[h]
\begin{center}
\includegraphics[width=15cm]{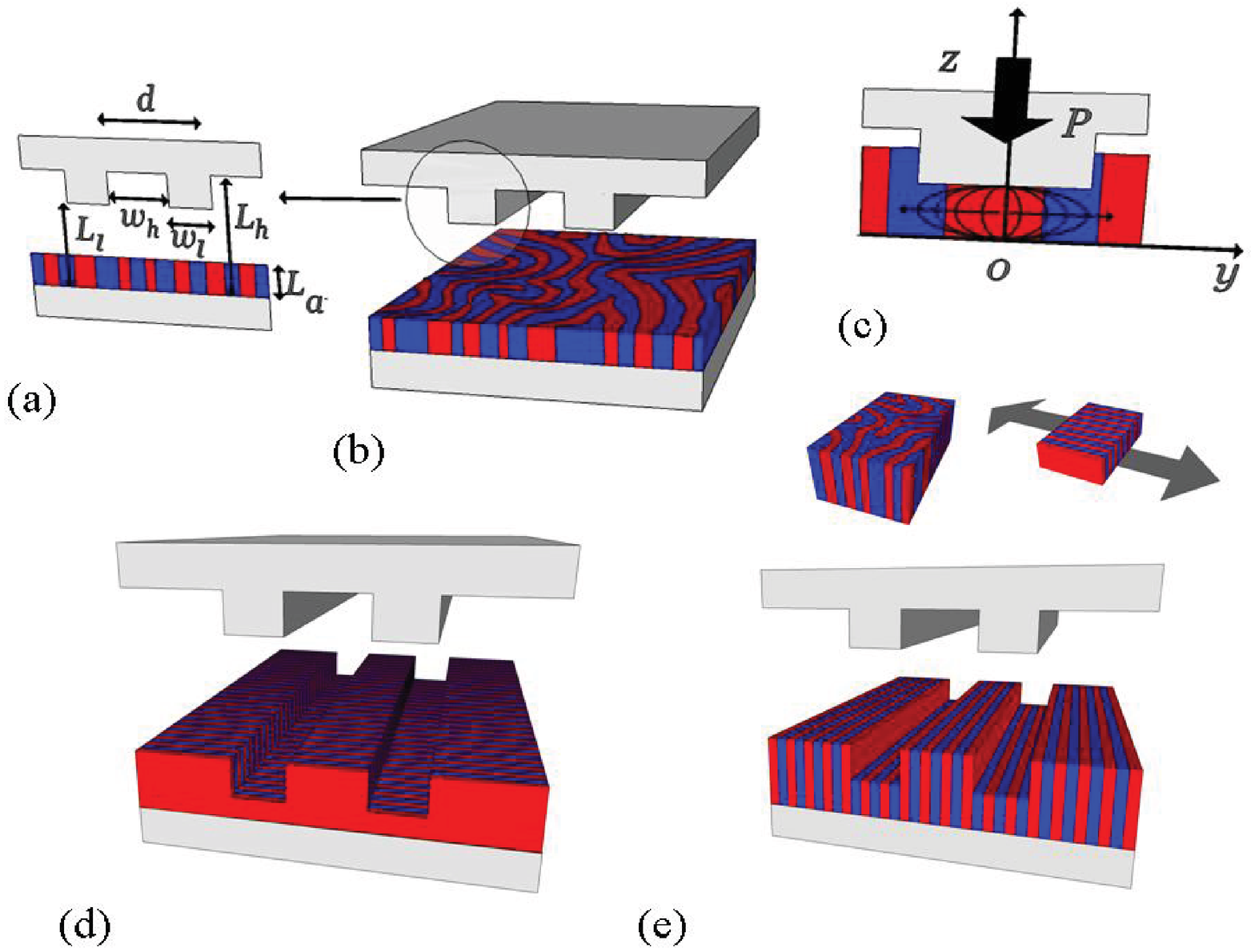}
\caption{Schematics of the NIL process. (a) The geometry of a teflon-coated mold made of silicon.
(b) The mold is pressed at a temperature above
$T_g$ against a BCP film in its lamellar phase pre-oriented perpendicular to a substrate. (c)
Nano-rheology in the film plays an important role at high shear rates obtained
for higher temperatures, and (d) the resulting
lamellae are oriented perpendicularly to both the bottom substrate and the grooves.
(e) When surface energy effects dominate, the lamellae are oriented perpendicular to the substrate and parallel to the grooves.
}
\label{fig1}
\end{center}
\end{figure}

Nano Imprint Lithography (NIL)  has been proposed in the mid-nineties\cite{Chou95} and consists in first embossing a molten
polymer film above its glass transition temperature $T_g$ with
a mold exhibiting micro-scale features (Fig. \ref{fig1}), which
is produced only once but can be re-used over 1000 times.
Then, the combined substrate/film/mold is cooled down below $T_g$ and the film is separated
from the weakly adhering fluorocarbon-coated mold.
NIL has been widely used  for homopolymers\cite{li00,jung04}.
However, BCP films were studied only in a few cases ~\cite{li04, kim08, voet11} and no attempt to organize defectless perpendicular structures over large distances (tens of micrometers) was reported.
The ease of implementation and low cost of the method present tremendous advantages as compared to other techniques such as 193 nm deep UV, electron or
focused ion beam lithography.\cite{booklan}

In this work, we demonstrate how nano-rheology of the confined BCP film ~\cite{ruiz07b, sundrani, angelescu} and surface interactions due to the mold
are used to control nano-structures, enabling
unprecedented defect-free film ordering on the wafer scale. Hence,
the flexibility of mold design provides an efficient top-down way to tailor BCP patterns transferrable onto a silicon substrate.
In particular, NIL  produces patterns in complex geometries that are hardly reachable using state-of-the-art fabrication methods,
and can lead to a wealth of applications for integrated high-density devices such as magnetic memories, crossbar circuits and metamaterials.

Symmetric diblock copolymers of  PS$_{\rm 52K}$-{\it b}-PMMA$_{\rm 52K}$  with polydispersity index of 1.09 were purchased
from {\it Polymer Source Inc} and exhibit a bulk lamellar phase of period $L_0 = 49 $\,nm~\cite{stoyko05}.
High-energy silicon wafers are cleaned and coated with a self-assembled monolayer (SAM) of octadecyltrichlorosilane.
A UV/ozone treatment of the SAM is used to modify the surface energy and to induce perpendicular orientation of the BCP lamellae
with respect to the substrate\cite{liu09}.
Finally, thin BCP films are spin-coated on the modified wafer, nanoimprinted and examined with SEM.

\begin{figure}[h]
\begin{center}
\includegraphics[width=10cm]{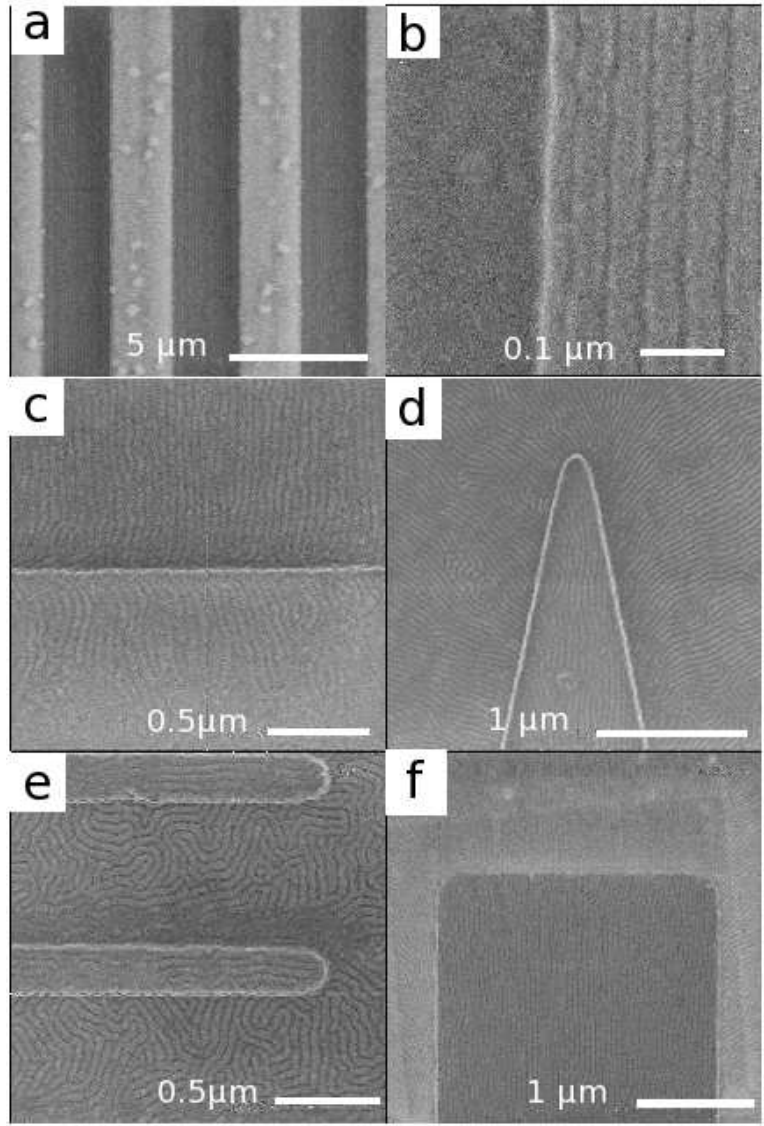}
\caption{Scanning electron microscopy (SEM) images of PS$_{\rm 52K}-b-$PMMA$_{\rm 52K}$ diblock copolymer films after NIL processing.
(a),(b) Processing at $T=120^\circ$ C with parallel alignment in the thick darker regions.
The 50\,nm lamellar period is clearly visible in (b).
(c)-(f) Processing at $T=170^\circ$ C. In (c) orientation perpendicular to the grooves is clearly visible (see also Fig. \ref{fig4}).
In (d) the perpendicular orientation of the lamellar phase around a tip suggests a similar orientation of flow streamlines. In (e) the small groove width leads to a large surface-energy effect and parallel orientation of the thickest (lighter) regions, whereas
random orientation (that, nevertheless, mainly starts perpendicular to the walls), is observed for the lower (darker) regions.
In (f) both surface effects and dominant flow from the upper part of the groove contribute to a perfect parallel orientation.}
\label{fig2}
\end{center}
\end{figure}

\begin{figure}[h]
\begin{center}
\includegraphics[width=14cm,bb=0 0 550 250]{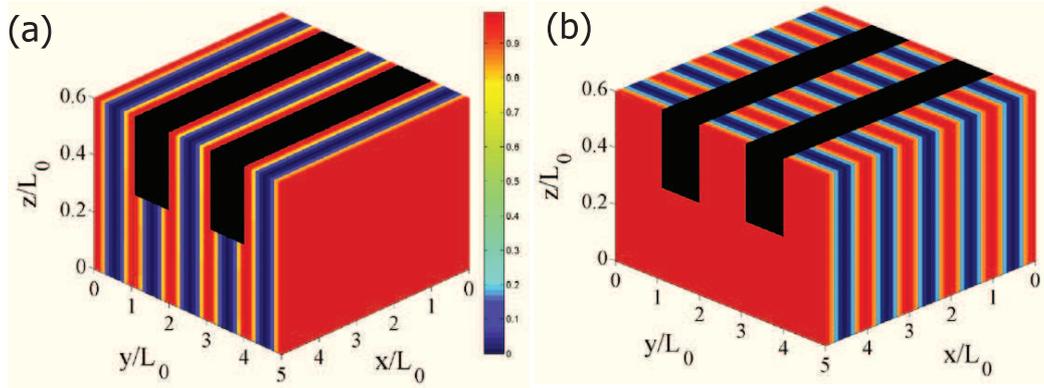}
\caption{
Three-dimensional SCFT simulations of a BCP film in a NIL setup. (a) Starting from a disordered system, the simulations converge to lamellae oriented parallel to the groove.
(b) Starting from a perpendicular orientation, the system stays in the this state, though with higher free energy.}
\label{fig3}
\end{center}
\end{figure}

For samples processed at low temperatures of about $T=110-120^\circ$\,C,  defect-free lamellae having a periodicity of $\approx L_0$ (Figs. \ref{fig2} a and b) are observed and are oriented parallel to the groove edges over tens of micrometers.
Self-consistent field theory (SCFT) calculations~\cite{xingkun10,xingkun} (see Supporting Information) in a three-dimensional geometry mimicking the experimental
substrate and mold (Fig. \ref{fig1}) support the experimental findings and are shown in Fig. \ref{fig3}.
This agreement between theory and experiment strengthens our claim that, in this low temperature case,  the observed overall orientation is determined by the surface energy.
Furthermore, if the substrate has no preference for one of the two blocks and the mold has a slight preference for one of them,
well-ordered lamellae oriented perpendicular to the substrate and parallel to the grooves (Figs.\ref{fig2}a, \ref{fig2}b and \ref{fig3}) are found.
Indeed, for thin enough grooves in the $y$ direction ($\omega_h \lesssim 30 L_0$), the interaction with
the mold side-walls prevails and is the main driving force for the  orientation of the lamellae.\cite{xingkun}

Surprisingly, for samples processed at higher temperatures of around 170$^\circ$\,C  the lamellae are
found to be {\it perpendicular} to the groove edges (Fig. \ref{fig2}c). This finding cannot be solely explained by
equilibrium film considerations.
The organization of BCP around a mold tip (Fig. \ref{fig2}d) suggests similar directed flow streamlines and points to
important rheological effects that have been previously noticed for cylinders oriented parallel to the substrate\cite{voet11}.

The rheology of BCP melts has attracted considerable attention since the 1970's due to its particular relevance
to extrusion~\cite{angelescu, hamley01}. Using the lubrication approximation (see Supporting Information),
the film flow is expressed in terms of the shear rate $\dot{\gamma}$, which is proportional to $P y L_A /(\eta\omega_l^2)$,
where $P$ is the applied pressure, $L_A$ the initial film thickness, $\eta$ the film viscosity, and $y$ the distance from the zero velocity plane (Fig. \ref{fig1}c).
The maximum shear rate is obtained at the groove walls $(y=\pm \omega_l/2, z=L_l)$  and ($y=\pm \omega_l/2,z=0$).
For low shear rates (less than  $\approx 1$\,s$^{-1}$),
it was shown both in experiments\cite{hamley01,chen} and in numerical simulations\cite{fraser06,lisal07}
that the lamella normals are parallel to the shear gradient; namely, the lamellae are expected to be parallel to the substrate.

\begin{figure}[h]
\begin{center}
\includegraphics[width=12cm]{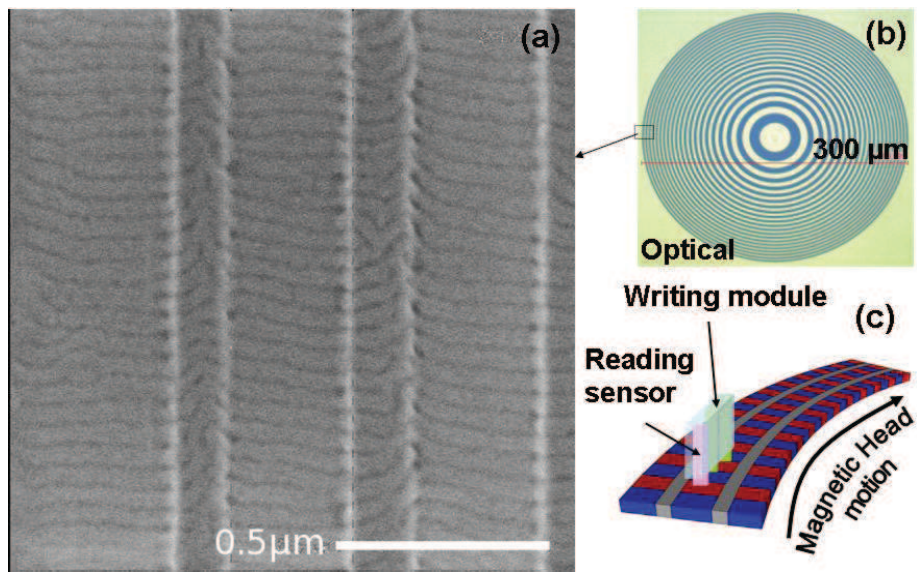}
\caption{
(a) SEM top-view picture of concentric circles imprinted in a PS-PMMA BCP film showing a perpendicular
organization of the BCP film with respect to the circular tracks.
(b) Optical microscopy picture at a much larger scale of 300\,$\mu$m.
(c) Schematic drawing showing the motion of a magnetic reading head in a magnetic storage disk.}
\label{fig4}
\end{center}
\end{figure}

However, for higher shear rates the orientation of the lamella normals is perpendicular both to the shear gradient and to the
flow direction, meaning
that the lamellae are perpendicular to both the substrate and groove walls.
With typical thickness and width values, $L_A$=50\,nm, $\omega_l$=500\,nm, and applied pressure $P=1.5 \times 10^6$\,Pa, we estimate the maximum shear rate to be
$\dot{\gamma}\approx 0.1$\,s$^{-1}$ at 110$^\circ$\,C (viscosity $\eta \approx 2 \times 10^7$\,Pa.s) and $\dot{\gamma}\gtrsim 100$\, s$^{-1}$ at 170$^\circ$\,C
($\eta \approx 2 \times 10^4$\,Pa.s).
This means that  a lamellar orientation perpendicular to the groove walls is expected at 170$^\circ$\,C,
as it is indeed observed (Figs. \ref{fig2}c, \ref{fig2}d and \ref{fig4})

Full three-dimensional SCFT calculations show that a lamellar orientation parallel to the grooves is obtained when the film initial condition is  in the disordered state (above ODT)  (Fig. \ref{fig3}a).
However, when  we take the lamellae to be oriented perpendicular to the grooves as an initial condition which mimicks the polymer flow, the system keeps this orientation although it has a higher free energy (Fig. \ref{fig3}b). This suggests that a perpendicular orientation is not eventually destroyed by surface energy but rather is a metastable state of the system.
We conclude that both film nano-rheology and surface energy are important in determining the BCP organization
as shown in (Fig. \ref{fig2}e). In this latter case, the small width of the grooves
implies a relatively large contribution to the surface energy of the vertical walls,
leading to parallel orientation inside the grooves,
whereas random orientation (with a preference to perpendicular orientation) is found close to the groove walls.

The competing effects of surface energy and nano-rheology provide a unique way to control and design
nano-structures on silicon wafers.
Large scale defect-free orientation of the lamellae can be reached by using a dedicated mold geometry to control the flow.
This can be  achieved by injecting polymer  with large shear rate into
the mold grooves in places
where an orientation of the lamellae perpendicular to the walls is desired, while
reducing the polymer flow in places where parallel orientation is needed. This is illustrated in Fig. \ref{fig2}f where the polymer flows from the lighter colored thinner regions to the central darker and thicker one. The upper downward flow reaches the thicker zone with a much higher shear rate than the lateral flow because the upper injection region has a much larger extension than the lateral ones (the shear rate increases with the $y$ coordinate, see Eq.(4) in SI).  Surface energy effects are designed to reinforce the flow effect since the lateral side walls  were made longer than the upper wall.

Flow is in fact particularly efficient for obtaining defect-free structures because of its long-range effect while surface energies represent a
short-range one.
Moreover, as the viscosity and shear rate can be varied over orders of magnitude
($\eta\approx 10^4$ to $10^7$ Pa.s.) by changing the temperature over $\sim 50^\circ$\,C only, both effects can be furthermore combined. In Fig. 2f  a high-temperature step causing defect healing was added to a low-temperature step used
for pre-ordering at the walls. Namely, after a first step of 2 minutes at $150^\circ$\,C for evaporating fully  the solvent, a pre-ordering step was performed at $120^\circ$\,C for 7 hours followed by a defect healing step at $170^\circ$\,C for 72 hours. A force of 10 kN was applied onto the 4 inches wafer.

Our NIL method provides a versatile approach to order BCP nanostructures in various geometries.  In Fig. \ref{fig4} we show a radial ordering which is induced by a flow in a system of cylindrical symmetry. A mold with concentric grooves features was created and is shown in Fig. \ref{fig4}b. The remarkable radial ordering of the lamellae was obtained by a high temperature step to favor long-range ordering by the flow.
This possibility of imprinting circular geometries is of great advantage for applications such as circular reading tracks for ultra-high density \it  bit patterned media \rm (BPM)
proposed for magnetic storage disks\cite{Lan}.
Our present scheme can indeed create ultra-dense circular features (the density is fully tunable by the polymer molecular weight) perpendicular to
a magnetic head motion (Fig. \ref{fig4}c). This is  fully compatible with actual rotating magnetic disk,
unlike other approaches that lead to patterns that are orthogonal in the $x$-$y$ directions.

In conclusion we have demonstrated a simple way to generate and tailor defect-free nanostructures at the 10\,nm
resolution. Long-range nano-rheology and shorter-range surface energy effects can be efficiently combined  to provide dedicated pattern geometries.  For further applications the BCP patterns processed by NIL will play the role of an intermediate mask or of a template. We have successfully checked  that such a BCP film can be transferred onto a silicon wafer by Reactive Ion Etching (see Supporting Information). In the future we foresee that metals can also be included into BCP structures using, for example, deposition of metallic layers onto pre-patterned templates\cite{Lu},
electrodeposition\cite{Thurn} and immersion in aqueous metal-salt solution\cite{Chai}. When processed by our cost-effective NIL method, hybrid ordered nanostructures could  provide a wealth of applications  for ultra-high
density magnetic recording media\cite{Thurn}, metal nano-structures for metamaterials and
plasmonic circuits, sensors\cite{Mist} and memories\cite{Hong}.

\section*{Experimental}
\it Surface Modification of  Silicon Substrates\rm :  Silicon wafers (p-type, 500 $\mu m$ thick, purchased from MEMC Elect. Materials) were cleaned by sonication in purified Millipore water and 2-propanol, followed by exposure to UV under an oxygen atmosphere for 30 min. The cleaned wafers were silanized in a 2mM solution of octadecyltrichlorosilane (OTS) in heptane for 1 day. The wafers were put onto a heat plate at $100^\circ$C for 5 min, then sonicated in toluene and dried with nitrogen before UV/ozone treatment.
\\
\it Surface Energy Control\rm : The SAM is subsequently oxidized to obtain the desired surface-energy value through UV/ozone treatment as previously shown in earlier work \cite{liu09}.
\\
\it Polymer Film Deposition \rm: 1 wt \% solution of diblock copolymer of PS52K-b-PMMA52K, (Polymerization index:1.09) of symmetric composition (purchased from Polymer Source Inc.) in toluene was spin coated at 1800 rpm onto the silanized silicon wafer, in order to produce a BCP films with thickness of about 40-50 nm. Subsequently, the samples were annealed in a vacuum oven with a pressure less than 30 mbar at $170^\circ$ C for 1 day.\\
\it NanoImprinting Procedure \rm :
Imprint experiments have been carried out using an EVG@520HE press under vacuum of about 0.1 mbar. The mold used was a 4in. silicon wafer with engraved nano-structures of 50 nm in height and between 250 nm and 8 $\mu$m in width. An overall pressure of 1.5 MPa was then applied to all samples at temperatures of $120^\circ$C (samples of Figs. 2a and 2b) or $170^\circ$C (samples of  Figs. 2c-2e) during 18h.
Critical Dimension Scanning Electron Microscopy (CD-SEM) were performed using a Hitachi 9300 instrument operating at low voltage (500 V).\\

\section*{Acknowledgements}
Funding from Chimtronique and Nanosciences programs of the CEA, the network for advanced research ``Triangle de la Physique'' (POMICO project), the U.S.-Israel Binational Science Foundation under Grant No.
2006/055, the Israel Science Foundation under Grant No. 231/08, and the Center for Nanoscience and Nanotechnology at Tel Aviv
University is gratefully acknowledged.


\bibliographystyle{advanced}
\bibliography{panghung}

\end{document}